\documentclass[12pt]{article}
\bigskip

\newcommand{\Si}[1]{\text{Si}\!\br{#1}}
\newcommand{\Ci}[1]{\text{Ci}\!\br{#1}}
\newcommand{\bSi}[1]{\text{Si}\bbr{#1}}
\newcommand{\bCi}[1]{\text{Ci}\bbr{#1}}

\usepackage{amsmath}
\usepackage{amsfonts}
\usepackage{amssymb}
\usepackage[bf]{caption}
\usepackage{cite}
\usepackage{psfrag,graphicx}
\usepackage{textcomp}
\usepackage{wrapfig}
\usepackage{upgreek}
\usepackage{hyperref}

\newcommand{\cte}[1]{``#1"}

\newcommand{\prom}{\textperthousand}
\newcommand{\pprom}{\textpertenthousand}

\newcommand{\ep}{\varepsilon}

\renewcommand{\j}{\text{j}}

\newcommand{\e}{\text{e}}

\newcommand{\br}[1]{\left({#1}\right)}
\newcommand{\bbr}[1]{\big({#1}\big)}

\newcommand{\brn}[1]{\!\left({#1}\right)} 
\newcommand{\bbrn}[1]{\!\big({#1}\big)}

\newcommand{\brh}[1]{\left[{#1}\right]}
\newcommand{\brc}[1]{\left\{{#1}\right\}}
\newcommand{\brb}[1]{\left|{#1}\right|}

\DeclareMathAlphabet{\mathsfbfsl}{T1}{cmss}{bx}{sl}

\newcommand{\dd}{\mbox{d}}

\begin{document}

\title{{\Large \bf The capacitance of the circular parallel plate capacitor obtained by solving the Love integral equation
using an analytic expansion of the kernel}}
\author{Martin Norgren and B. L. G. Jonsson}
\maketitle

\begin{center}
Division of Electromagnetic Engineering\\ Royal Institute of
Technology\\ SE-100 44 Stockholm, Sweden \\[6mm]
Corresponding author: Martin Norgren\\
Email: martin.norgren@ee.kth.se\\ Tel: +46 8 7907410; Fax: +46 8
205268\\[10mm]
\end{center}

\newpage

\begin{abstract}
The capacitance of the circular parallel plate capacitor is
calculated by expanding the solution to the Love integral equation
into a Fourier cosine series. Previously, this kind of expansion has
been carried out numerically, resulting in accuracy problems at
small plate separations. We show that this bottleneck can be
alleviated, by calculating all expansion integrals analytically in
terms of the Sine and Cosine integrals. Hence, we can, in the
approximation of the kernel, use considerably larger matrices,
resulting in improved numerical accuracy for the capacitance. In
order to improve the accuracy at the smallest separations, we
develop a heuristic extrapolation scheme that takes into account the
convergence properties of the algorithm. Our results are compared
with other numerical results from the literature and with the
Kirchhoff result. Error estimates are presented, from which we
conclude that our results is a substantial improvement compared with
earlier numerical results.
\end{abstract}

\section{Introduction}

The exact capacitance of the circular parallel plate capacitor, with
infinitely thin plates, remains an unsolved problem in potential
theory, in the sense that to this date no explicit analytical
solution has been reported. However, the problem can be formulated
as a Fredholm integral equation of the second kind, known as Love's
integral equation \cite{Love}, which can be solved numerically.

To our knowledge, the up to date most accurate studies of the
capacitance at small plate separations are the ones by Wintle and
Kurylowicz \cite{Wintle_Kurylowicz} and by Carlson and Illman
\cite{Carlson_Illman}. Both of these studies have been used as
benchmarks for solutions obtained by other methods; see e.g.
\cite{Nishiyama+Nakamura,Donolato,Hwang+Given}. Wintle and
Kurylowicz \cite{Wintle_Kurylowicz} use an El-Gendi method
\cite{ElGendi} to rewrite the Love equation and apply numerical
integration using the Clenshow-Curtis quadrature method
\cite{Clenshav+Curtis} to obtain the capacitance. Carlson and Illman
\cite{Carlson_Illman} solve the Love equation through an expansion
of the kernel into a Fourier-cosine series. Later, they have also
extended that method to solve the three-plate problem by means of
coupled Love type equations \cite{Carlson_Illman2}. It is known
\cite{Love} that for small plate separations, a solution obtained
via a series expansion of the kernel converges slowly, requiring a
large number of expansion terms. To calculate the expansion
coefficients of the kernel, Carlson and Illman \cite{Carlson_Illman}
use numerical integration. Hence, their method is limited by a
combination of the accuracy of the integration and the large number
of terms needed. The accumulated errors effectively limit the
expansion to about 100 terms, which is insufficient for convergence
at very small separations.

In this paper, we show that all of the integrals in the series
expansion in \cite{Carlson_Illman} can be expressed analytically in
terms of the well-studied Sine- and Cosine integrals. In this way,
we improve the numerical accuracy of the expansion coefficients up
to the accuracy of the evaluations of the Sine- and Cosine
integrals, which makes it possible to increase the number of
expansion functions considerably. Hence, in our method, the
numerical accuracy in the capacitance is mainly limited by the
truncation of the number of expansion functions. Thus, we can
present improved results for the capacitance at small plate
separations; results that surpass the results in
\cite{Wintle_Kurylowicz,Carlson_Illman}, both in correctness and in
the significant numbers of digits. Our results are also in excellent
agreement with the result by Kirchhoff\cite{Kirchhoff}, which
becomes increasingly accurate when the plate separation tends to
zero \cite{Hutson,Sneddon}.

The paper is organized as follows: In Section \ref{avsnitt:problem},
we review the expansion method, originally presented in
\cite{Carlson_Illman}. In Section \ref{avsnitt:integraler}, we
derive the analytical expressions for all the expansion integrals.
Numerical results are presented in Section \ref{avsnitt:numeriska}
and Section \ref{avsnitt:diskussion} contains some conclusions.

\section{Problem formulation and initial analysis}

\label{avsnitt:problem}

\begin{wrapfigure}{r}{50mm}
\centering \psfrag{a}{$a$} \psfrag{d}{$d$}
\includegraphics{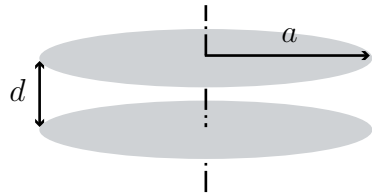}
\caption{The circular parallel plate capacitor (side view slightly
from above).} \label{fig:konding}
\end{wrapfigure}

The circular parallel plate capacitor is depicted in Figure
\ref{fig:konding}. The distance between the circular plates is
denoted $d$ and their common radius is denoted $a$. The model is
idealized in the sense that the plates have zero thicknesses.
Following the notation used in many of the previous studies of this
problem, we let $\kappa=d/a$ denote the normalized separation
between the plates.

The capacitance of the parallel plate capacitor is
\cite{Carlson_Illman}
\begin{align}
C = 4\ep_0 a \int_0^1 f\brn{s} \dd s, \label{eq:PF:3}
\end{align}
where the function $f\brn{s}$ is the solution to the modified Love
integral equation
\begin{align}
f\brn{s} - \int_0^1 K\brn{s,t} f\brn{t}\dd t = 1, \qquad 0\leq  s
\leq 1, \label{eq:PF:1}
\end{align}
with kernel
\begin{align}
K\brn{s,t} = \frac{\kappa}{\pi}\brh{\frac{1}{ \kappa^2+\br{s-t}^2} +
\frac{1}{\kappa^2+\br{s+t}^2}}.  \label{eq:PF:2}
\end{align}
Note that in the original derivation by Love \cite{Love}, the kernel
and the function are defined in the range $-1 \leq s,t \leq 1$, and
the kernel has only one term, but since $f\brn{s}$ can be shown to
be even one can instead use the formulations (\ref{eq:PF:1}) and
(\ref{eq:PF:2}) (an elegant and short derivation of Love's integral
equation can be found in \cite{Sneddon}).

To solve Equation (\ref{eq:PF:1}) numerically, we follow the
approach in \cite{Carlson_Illman} and expand the kernel and the
unknown function into the Fourier-cosine expansion functions
\begin{align}
\psi_m\brn{s} = \sqrt{2-\updelta_{m0}} \cos\brn{m\pi s}, \qquad
m=0,1,\ldots, \label{eq:PF:4}
\end{align}
which in our study have been normalized to fulfil the orthogonality
relation
\begin{align}
\int_0^1 \psi_m\brn{s} \psi_{m^\prime}\brn{s} \dd s =
\updelta_{mm^\prime},
\end{align}
where $\updelta_{mm^\prime}$ denotes the Kronecker delta function.
Note that similarly to $f\brn{s}$ all $\psi_m$-functions have
vanishing first derivatives at $s=0$.

Carrying out the expansions of $f\brn{s}$ and $K\brn{s,t}$, in terms
of $\brc{\psi_m}$, we obtain
\begin{align}
& f\brn{s} = \sum_{m=0}^\infty f_m \psi_m\brn{s}, \label{eq:PF:5} \\
& f_m = \int_0^1 f\brn{s} \psi_m\brn{s}\dd s, \label{eq:PF:6} \\
& K\brn{s,t} = \sum_{m=0}^\infty \sum_{n=0}^\infty K_{mn}
\psi_m\brn{s} \psi_n\brn{t}, \label{eq:PF:7} \\
& K_{mn} = \int_0^1 \int_0^1 K\brn{s,t} \psi_m\brn{s} \psi_n\brn{t}
\dd s \dd t,  \label{eq:PF:8}
\end{align}
which yield the following infinite linear system of equations for
the coefficients $\brc{f_n}_{n=0}^\infty$:
\begin{align}
\sum_{n=0}^\infty \br{\updelta_{mn} - K_{mn}} f_n = \updelta_{m0},
\qquad m=0,1,\ldots. \label{eq:PF:9}
\end{align}
From (\ref{eq:PF:3}), (\ref{eq:PF:5}) and the orthogonality of
(\ref{eq:PF:4}), the capacitance reduces to
\begin{align}
C = 4\ep_0 a f_0, \label{eq:PF:10}
\end{align}
where $f_0$ is simply the (0,0)-element in the inverse of the matrix
with elements $\updelta_{mn} - K_{mn}$.

In the numerical implementation the matrix is truncated into the
size $0\leq m,n \leq N$, where we call $N$ the truncation number. It
is well-known \cite{Hutson,Carlson_Illman} that at small
separations, $\kappa$, large values of $N$ are required to obtain
convergence of $f_0$ in (\ref{eq:PF:10}). Also, for the accuracy of
the result, it is crucial that the matrix elements $K_{mn}$, given
by the integrals in (\ref{eq:PF:8}), have been calculated with a
high accuracy.

For small values of $\kappa$ the kernel $K\brn{s,t}$ has a
pronounced crest at $s=t$, making it difficult to evaluate the
integrals in (\ref{eq:PF:8}) numerically. In \cite{Carlson_Illman}
this problem has been alleviated, by adding and subtracting a
suitable term to the kernel, thereby removing the crest in one of
the integrals, and making the inner integral in the other integral
available for explicit evaluation. A related procedure has also been
used earlier in \cite{Wintle_Kurylowicz}. However, a remaining
problem is that for large values of $m,n$ the expansion functions
(\ref{eq:PF:4}) yield rapidly oscillating integrands, which (when
integrated numerically) result in slow convergence and poor
accuracy. Hence, it would be beneficial if all integrals,
encountered when expanding the kernel, could be expressed
analytically in terms of established functions, readily available
for numerical evaluation with a high accuracy. In the next section
we will show that this is indeed the case.

\section{Expansion of the kernel into Sine and Cosine integrals}

\label{avsnitt:integraler}

In this section, we derive the analytical expressions for the
expansion of the kernel $K\brn{s,t}$. First, we notice from
(\ref{eq:PF:2}) and (\ref{eq:PF:8}) that the expansion coefficients
have the property $K_{nm}=K_{mn}$, resulting in a symmetric matrix.
For the special cases when the indices coincide and/or becomes zero,
it is advantageous for the numerical evaluation to derive special
simplified expressions for $K_{mn}$. In most of the derivations, we
utilize the following properties of the Sine and Cosine integrals
\cite{Abramowitz+Stegun}:
\begin{align}
& \Si{z^\ast} = \brh{\Si{z}}^\ast, \label{eq:Si:conj}\\
& \Ci{z^\ast} = \brh{\Ci{z}}^\ast, \label{eq:Ci:conj}\\
& \Si{-z} = -\Si{z}, \label{eq:Si:sign}\\
& \Ci{-z} = \Ci{z} - \j\pi  \quad \br{0 < \arg{z} < \pi},
\label{eq:Ci:sign}
\end{align}
where $\ast$ denotes complex conjugation and $\j$ denotes the
imaginary unit.

\subsection{Case $m=n=0$}

By elementary integrals, we obtain
\begin{align}
K_{0,0} & = \int_0^1  \int_0^1 K\brn{s,t}\dd s \: \dd t =
\frac{\kappa}{\pi}\int_0^1 \brc{\int_0^1
\brh{\frac{1}{\kappa^2+\br{s-t}^2} +
\frac{1}{\kappa^2+\br{s+t}^2}} \dd s} \dd t \nonumber \\
& = \frac{1}{\pi}\int_0^1 \brh{\arctan\brn{\frac{t+1}{\kappa}} -
\arctan\brn{\frac{t-1}{\kappa}}} \dd t \nonumber \\
& = \frac{1}{2\pi}\brh{4\arctan\brn{\frac{2}{\kappa}} - \kappa
\ln\brn{1+\frac{4}{\kappa^2}}}. \label{eq:int:K00}
\end{align}

\subsection{Cases $m=0, n > 0$}

Here the inner integral is the same as in the previous case,
resulting in
\begin{align}
K_{0n} & = \sqrt{2} \int_0^1  \int_0^1 K\brn{s,t} \cos\brn{n\pi t}
\dd
s \: \dd t \nonumber \\
& = \frac{\sqrt{2}}{\pi} \int_0^1
\brh{\arctan\brn{\frac{t+1}{\kappa}} -
\arctan\brn{\frac{t-1}{\kappa}}} \cos\brn{n\pi t} \dd t.
\label{eq:int:K0n:1}
\end{align}
Defining the function
\begin{align}
I_1\brn{ \beta, \kappa, \alpha} = \int_0^1 \cos\brn{\beta t}
\arctan\brn{\frac{t+\alpha}{\kappa}} \dd t, \label{eq:int:K0n:2}
\end{align}
it follows that
\begin{align}
K_{0n}  = \frac{\sqrt{2}}{\pi} \brh{I_1\brn{n\pi, \kappa, 1} -
I_1\brn{n\pi, \kappa, -1}}.  \label{eq:int:K0n:3}
\end{align}
By a change of variable and integration by parts, we obtain
\begin{align}
I_1\brn{\beta,\kappa,\alpha} & = \int_0^1 \cos\brn{\beta t}
\arctan\brn{\frac{t+\alpha}{\kappa}} \dd t =
\brc{u=\frac{t+\alpha}{\kappa}} \nonumber \\ & = \kappa
\int_{\alpha/\kappa}^{\br{1+\alpha}/\kappa}
\cos\bbrn{\beta\brn{\kappa u-\alpha}} \arctan\brn{u}  \dd u \nonumber \\
& =
\frac{\sin\brn{\beta}}{\beta}\arctan\brn{\frac{1+\alpha}{\kappa}} -
\frac{1}{\beta} \int_{\alpha/\kappa}^{\br{1+\alpha}/\kappa}
\frac{\sin\bbrn{\beta\brn{\kappa u-\alpha}}}{1+u^2}\dd u \nonumber \\
& =
\frac{\sin\brn{\beta}}{\beta}\arctan\brn{\frac{1+\alpha}{\kappa}}
\nonumber \\
& \quad\;\; + \frac{1}{\beta} \text{Im}\left\{
\sin\bbrn{\beta\br{\alpha+\j\kappa}} \Big( \bCi{\beta\br{\alpha +
\j\kappa }} - \bCi{\beta\br{\alpha + 1 + \j\kappa }} \Big) \right. \nonumber \\
& \qquad\quad\quad - \left.
\cos\bbrn{\beta\br{\alpha+\j\kappa}}\Big(\bSi{\beta\br{\alpha +
\j\kappa }} - \bSi{\beta\br{\alpha + 1 + \j\kappa }} \Big) \right\}.
\label{eq:int:K0n:4}
\end{align}
The evaluation of the last integral was carried out in the Maple
software, and the result was simplified using the
trigonometric-hyberbolic addition formulas and the properties
(\ref{eq:Si:conj}) and (\ref{eq:Ci:conj}).

Insertion of (\ref{eq:int:K0n:4}) into (\ref{eq:int:K0n:3}) and
further simplifications yield
\begin{align}
K_{0n} = \frac{\sqrt{2}}{n \pi^2} &\text{Im}\Big\{
\cos\bbrn{n\pi\br{1+\j\kappa}} \bSi{n\pi\br{2+\j\kappa}} +
\cos\bbrn{n\pi\br{1-\j\kappa}} \Si{-\j n\pi \kappa} \nonumber \\
& \qquad -\sin\bbrn{n\pi\br{1+\j\kappa}} \bCi{n\pi\br{2+\j\kappa}}
-\sin\bbrn{n\pi\br{1-\j\kappa}} \Ci{-\j n\pi \kappa}
 \Big\}.
\end{align}

\subsection{Cases $m \neq n, m > 0, n > 0$}

Here, we obtain
\begin{align}
K_{mn} & = 2 \int_0^1  \int_0^1 K\brn{s,t} \cos\brn{n\pi t}
\cos\brn{m\pi s} \dd t \: \dd s \nonumber \\
& = \frac{2}{\pi} \int_0^1 \int_0^1
\brh{\frac{\kappa}{\kappa^2+\br{s-t}^2} +
\frac{\kappa}{\kappa^2+\br{s+t}^2}} \cos\brn{n\pi t} \cos\brn{m\pi
s} \dd t \: \dd s. \label{eq:int:Kmn:1}
\end{align}
Defining the function
\begin{align}
I_2\brn{ \beta, \kappa, \alpha} = \int_0^1 \frac{\kappa
\cos\brn{\beta t}}{\kappa^2+\br{t+\alpha}^2} \dd t,
\label{eq:int:Kmn:2}
\end{align}
it follows that
\begin{align}
K_{mn}  = \frac{2}{\pi} \int_0^1 \brh{ I_2\brn{ n\pi, \kappa, -s} +
I_2\brn{ n\pi, \kappa, s} }\cos\brn{m\pi s} \dd s.
\label{eq:int:Kmn:3}
\end{align}
Again, using Maple and some simplifications it follows that
\begin{align}
I_2\brn{ \beta, \kappa, \alpha} & = \text{Im}\left\{
\sin\bbrn{\beta\br{\alpha+\j\kappa}} \Big( \bSi{\beta\br{\alpha +
\j\kappa }} - \bSi{\beta\br{\alpha + 1 + \j\kappa }} \Big) \right. \nonumber \\
&  \quad\;\;\: + \left. \cos\bbrn{\beta\br{\alpha+\j\kappa}} \Big(
\bCi{\beta\br{\alpha + \j\kappa }} - \bCi{\beta\br{\alpha + 1 +
\j\kappa }} \Big) \right\}.  \label{eq:int:Kmn:4}
\end{align}
Using (\ref{eq:Si:conj})-(\ref{eq:Ci:sign}), it follows that
\begin{align}
I_2\brn{ \beta, \kappa, -s} + I_2\brn{ \beta, \kappa, s} =
-\text{Im} & \Big\{ \; \sin\bbrn{\beta\br{s+\j\kappa}}
\bSi{\beta\br{s+1+\j\kappa}}
\nonumber \\
& + \cos\bbrn{\beta\br{s+\j\kappa}} \bCi{\beta\br{s+1+\j\kappa}}
\nonumber \\[2mm]
& + \sin\bbrn{\beta\br{s-\j\kappa}} \bSi{\beta\br{s-1-\j\kappa}}
\nonumber \\
& + \cos\bbrn{\beta\br{s-\j\kappa}} \bCi{\beta\br{s-1-\j\kappa}}
\Big\}. \label{eq:int:Kmn:5}
\end{align}
Inserting (\ref{eq:int:Kmn:5}) into the expression
(\ref{eq:int:Kmn:3}) for $K_{mn}$, we can write
\begin{align}
K_{mn} = \frac{2}{\pi} I_3\brn{n\pi, \kappa, m\pi},
\label{eq:int:Kmn:6}
\end{align}
where the function
\begin{align}
I_3\brn{\beta,\kappa,\gamma} & = \int_0^1 \big[ I_2\brn{ \beta,
\kappa, -s} +
I_2\brn{ \beta, \kappa, s} \big] \cos\brn{\gamma s} \dd s =
-\frac{1}{2}\text{Im}  \bigg\{ \int_0^1 \nonumber \\
&  \qquad\;\; \bigg[ \Big(\sin\bbrn{\br{\beta+\gamma}s+\j\kappa} +
\sin\bbrn{\br{\beta-\gamma}s+\j\kappa}\Big)
\bSi{\beta\br{s+1+\j\kappa}}
\nonumber \\[2mm]
& \qquad + \Big(\cos\bbrn{\br{\beta+\gamma}s+\j\kappa} +
\cos\bbrn{\br{\beta-\gamma}s+\j\kappa}\Big)
\bCi{\beta\br{s+1+\j\kappa}}
\nonumber \\[2mm]
& \qquad + \Big(\sin\bbrn{\br{\beta+\gamma}s-\j\kappa} +
\sin\bbrn{\br{\beta-\gamma}s-\j\kappa}\Big)
\bSi{\beta\br{s-1-\j\kappa}}
\nonumber \\[-1mm]
& \qquad + \Big(\cos\bbrn{\br{\beta+\gamma}s-\j\kappa} +
\cos\bbrn{\br{\beta-\gamma}s-\j\kappa}\Big)
\bCi{\beta\br{s-1-\j\kappa}} \bigg]
\dd s\bigg\} \nonumber \\
& = -\frac{1}{2}\text{Im}\big\{ I_4\brn{\beta+\gamma, \beta, \j
\kappa, 1+\j\kappa} + I_4\brn{\beta+\gamma, \beta, -\j \kappa,
-1-\j\kappa}
\nonumber \\
& \qquad\quad\:\: + I_4\brn{\beta-\gamma, \beta, \j \kappa,
1+\j\kappa} + I_4\brn{\beta-\gamma, \beta, -\j \kappa, -1-\j\kappa}
\big\}, \label{eq:int:Kmn:7}
\end{align}
and where
\begin{align}
I_4\brn{q,\beta,z_1,z_2} & = \int_0^1 \brh{\sin\brn{qs + \beta
z_1}\Si{\beta s + \beta z_2} + \cos\brn{qs + \beta z_1}\Ci{\beta s +
\beta z_2}}\dd s \nonumber \\
& = \frac{1}{q}\bigg[ \sin\brn{\beta z_1 + q}\bCi{\beta\br{1+z_2}} -
\cos\brn{\beta z_1 + q}\bSi{\beta\br{1+z_2}} \nonumber \\
& \qquad -\sin\brn{\beta z_1}\Ci{\beta z_2} + \cos\brn{\beta
z_1}\Si{\beta
z_2} \nonumber \\[2mm]
& \qquad +\cos\brn{\beta z_1 - q z_2}\Big(\bSi{\br{\beta
-q}\br{1+z_2}} -
\bSi{\br{\beta -q}z_2}\Big) \nonumber \\[2mm]
& \qquad -\sin\brn{\beta z_1 - q z_2}\Big(\bCi{\br{\beta
-q}\br{1+z_2}} - \bCi{\br{\beta -q}z_2}\Big) \bigg].
\label{eq:int:Kmn:8}
\end{align}
To derive (\ref{eq:int:Kmn:8}), we utilized the trigonometric
addition formulas and the integration formulas 5.31 and 5.32 in
\cite{Gradshteyn_Ryzhik}. Summarizing, we obtain
\begin{align}
K_{mn} = -\frac{1}{\pi} & \text{Im}  \big\{  I_4\bbr{\br{n+m}\pi,
n\pi, \j \kappa, 1+\j\kappa} + I_4\bbr{\br{n+m}\pi, n\pi, -\j
\kappa, -1-\j\kappa}
\nonumber \\
& \;\: + I_4\bbr{\br{n-m}\pi, n\pi, \j \kappa, 1+\j\kappa} +
I_4\bbr{\br{n-m}\pi, n\pi, -\j \kappa, -1-\j\kappa} \big\}.
\label{eq:int:Kmn:9}
\end{align}

\subsection{Cases $m = n > 0$}

In cases when $m=n>0$ the last two terms in (\ref{eq:int:Kmn:9}) are
not suitable for numerical evaluation, since in the final expression
in (\ref{eq:int:Kmn:8}) the limit $q\rightarrow 0$ must be taken.
Alternatively, we can start with setting $q=0$ in the integrand in
(\ref{eq:int:Kmn:8}); the result reduces to
\begin{align}
K_{nn} & = -\frac{1}{\pi} \text{Im}\bigg\{  I_4\brn{2n\pi,n\pi,\j
\kappa,
1+\j\kappa} + I_4\brn{2n\pi,n\pi,-\j \kappa, -1-\j\kappa} \nonumber \\
& \qquad + \sin\brn{\j n\pi\kappa} \int_0^1  \Big(
\bSi{n\pi\br{s+1+\j\kappa}} -
\bSi{n\pi\br{s-1-\j\kappa}}\Big)\dd s \nonumber \\
& \qquad + \cos\brn{\j n\pi\kappa} \int_0^1\Big(
\bCi{n\pi\br{s+1+\j\kappa}} -
\bCi{n\pi\br{s-1-\j\kappa}} \Big) \dd s \bigg\} \nonumber \\
& = -\frac{1}{\pi} \text{Im}\bigg\{  I_4\brn{2n\pi,n\pi,\j \kappa,
1+\j\kappa} + I_4\brn{2n\pi,n\pi,-\j \kappa, -1-\j\kappa} \nonumber \\
& \qquad\qquad + \j\sinh\brn{n\pi \kappa}\Big(
\br{2+\j\kappa}\bSi{n\pi\br{2+\j\kappa}} -\j\kappa
\Si{\j n\pi\kappa} \Big) \nonumber \\
& \qquad\qquad + \cosh\brn{n\pi\kappa}\Big(
\br{2+\j\kappa}\bCi{n\pi\br{2+\j\kappa}} -\j\kappa \Ci{\j
n\pi\kappa} -\j\pi \Big) \bigg\}.
\end{align}

\section{Numerical calculations}

\label{avsnitt:numeriska}

The here presented improved expansion is most useful at small
separations, wherefore we restrict our study to values $\kappa \leq
0.01$; results obtained at larger separations are readily available
in the literature, with some of the most accurate in
\cite{Wintle_Kurylowicz,Carlson_Illman}.

The value of the truncation number $N$ is mainly dictated by the
computational resources at hand. Here, we have used a maximum value
of $N=15000$, at the smallest separations.

All our results will be presented in terms of the normalized
capacitance $\mathcal{C}=C/\br{4\ep_0 a}$, where $4\ep_0 a$ is the
capacitance between two infinitely separated disks. Hence, it
follows from (\ref{eq:PF:10}) that $\mathcal{C}=f_0$.

\subsection{Extrapolation schemes}

To improve our numerical results, we have employed extrapolation.

\subsubsection{Power law model}

First, we considered a simple power law model for the capacitance:
\begin{align}
f_0\brn{N} = \hat{\mathcal{C}} - \beta
\br{N\kappa}^{-\alpha},\label{eq:NR:1}
\end{align}
where $f_0\brn{N}$ is the result obtained when using the truncation
$N$ in Equation (\ref{eq:PF:9}), $\hat{\mathcal{C}}$ is the
extrapolated estimate in the limit $N\rightarrow\infty$; $\alpha$
and $\beta$ (assuming $\alpha > 0, \beta >0$) are coefficients that
are determined together with $\hat{\mathcal{C}}$.

\begin{table}[h]
\centering
\begin{tabular}{|l||r|r|c|c||r|r|c|c|}
  \hline
&  \multicolumn{4}{c||}{$N\kappa=0.2$}  & \multicolumn{4}{c|}{$N\kappa=1$} \\
\cline{2-9}
 $\quad\kappa$  & $f_0\brn{N}\;$ & $\hat{\mathcal{C}}\quad\;$
 & $\alpha$ & $\beta\!\cdot\!10^2$  & $f_0\brn{N}\;\;$ &
 $\hat{\mathcal{C}}\quad\;\;$ & $\alpha$ & $\beta\!\cdot\! 10^3$ \\ \hline\hline
0.01    & 80.235    & 80.539   & 1.380 & 3.30 & 80.4312   & 80.4363   & 2.358 & 5.15 \\
0.005   & 158.937   & 159.240  & 1.431 & 3.03 & 159.1384  & 159.1436  & 2.377 & 5.14 \\
0.002   & 394.778   & 395.078  & 1.472 & 2.81 & 394.9827  & 394.9878  & 2.384 & 5.16 \\
0.001   & 787.647   & 787.946  & 1.486 & 2.73 & 787.8533  & 787.8585  & 2.386 & 5.16 \\
0.0005  & 1573.217  & 1573.516 & 1.492 & 2.71 & 1573.4238 & 1573.4290 & 2.388 & 5.16 \\
0.0002  & 3929.640  & 3929.849 & 1.496 & 2.69 & 3929.8466 & 3929.8518 & 2.389 & 5.16 \\
0.0001  & 7856.804  & 7857.102 & 1.498 & 2.69 & 7857.0105 & 7857.0156 & 2.391 & 5.15 \\
\hline
\end{tabular}
\caption{The extrapolation parameters, $\alpha, \beta$ and
$\hat{\mathcal{C}}$, obtained when fitting the power law
extrapolation model (\ref{eq:NR:1}) to $f_0\brn{N}, f_0\brn{N/2}$
and $f_0\brn{N/3}$. The estimate of the true capacitance is
$C=4\ep_0 a \hat{\mathcal{C}}$} \label{tabell:interpolering}
\end{table}

In our initial tests, the extrapolation parameters, $\alpha, \beta,
\hat{\mathcal{C}}$, were determined by fitting (\ref{eq:NR:1}) to
$f_0\brn{N}, f_0\brn{N/2}$ and $f_0\brn{N/3}$ (using rounded values
for the fractions of $N$ when necessary). In Table
\ref{tabell:interpolering} we present our results, for various
separations $\kappa$, for two different values of the product $N
\kappa$.

An important observation in Table \ref{tabell:interpolering} is that
for constant values of $N\kappa$ the values of the parameters
$\alpha$ and $\beta$ are approximately constant, regardless of the
value of $\kappa$. This is especially true for $N\kappa=1$, but also
for the smaller $\kappa$-values when
 $N\kappa=0.2$. The results for $N\kappa=0.2$ (farther from convergence) and
$N\kappa=1$ (closer to convergence) together indicate that the
extrapolation scheme over-estimates the capacitance. Another
observation from the data in Table \ref{tabell:interpolering} is
that for each value of $N\kappa$ the amount of extrapolation is
approximately the same, regardless of the value of $\kappa$.

\subsubsection{Heuristic model for improvement at low accuracies}

\label{avsnitt:numeriska:extrapol2}

\begin{figure}[h] \centering
\psfrag{gk}{$\times 10000$} \psfrag{gd}{$\times 10$}
\psfrag{p}{$\br{N\kappa}^{-1}$} \psfrag{C}{$h\brn{N\kappa}$}
\psfrag{k1}{$\kappa=0.01$} \psfrag{k2}{$\kappa=0.001$}
\psfrag{k3}{$\kappa=0.0001$}
\includegraphics{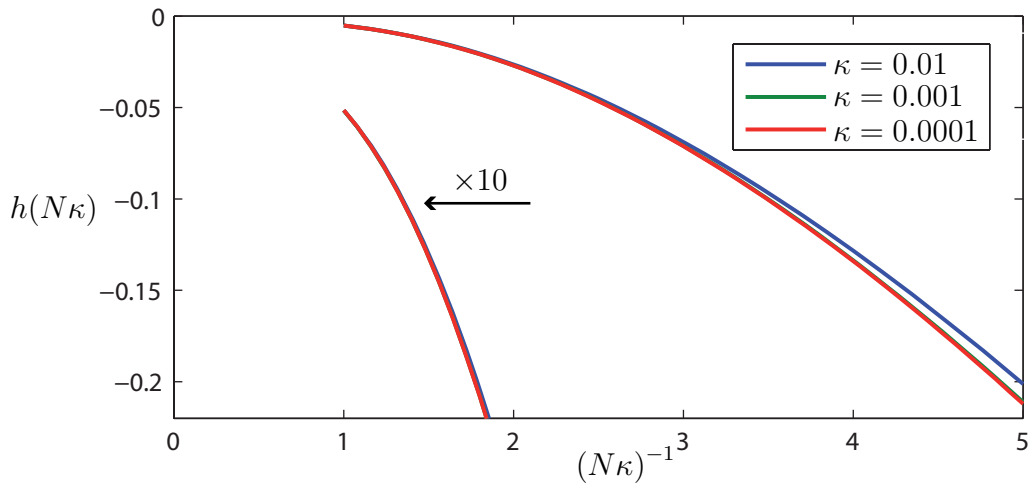}
\caption{Comparison of the estimated $h$-functions, obtained from
Equation (\ref{eq:NR:2}), for three different orders of the
separation $\kappa$. The $h$-function essentially indicates how
close to convergence the calculation of $C$ is, flatness of the
curve for high $N\kappa$ values indicates that $C$ has essentially
the distance $\brb{h\brn{Nk}}$ left to its approximately true value.
} \label{fig:hnp}
\end{figure}

Considering decreasing separations $\kappa$, the limitation of the
truncation $N$ will eventually force us to use smaller values of
$N\kappa$, which will push us farther from convergence. However, the
properties of the convergence, as demonstrated in Table
\ref{tabell:interpolering}, offer an opportunity to improve poorly
converged results, by using the following heuristic model:
\begin{align}
f_0\brn{N} = \tilde{\mathcal{C}} + h\brn{N\kappa},\label{eq:NR:2}
\end{align}
where the term $-\beta\br{N\kappa}^{-\alpha}$ in (\ref{eq:NR:1}) has
been replaced by a general function $h\brn{N\kappa}$. Using Equation
(\ref{eq:NR:2}), we have developed a heuristic extrapolation method
that is based on the assumption that when going from one
$\kappa$-value to the next smaller value the function $h$ remains
almost unchanged. To verify this, we have compared the $h$-functions
obtained at three different values of $\kappa$. We used equation
(\ref{eq:NR:2}), setting $\tilde{\mathcal{C}}=\hat{\mathcal{C}}$,
with $\hat{\mathcal{C}}$ taken from the seventh column in Table
\ref{tabell:interpolering}. The results are given in Figure
\ref{fig:hnp}, as function of $\br{N\kappa}^{-1}$ for the values
between the ones in Table \ref{tabell:interpolering}. The
$h$-functions in Figure \ref{fig:hnp} essentially illustrate the
convergence, with higher degree of convergence at lower
$\br{N\kappa}^{-1}$ values, where the $h$-values at
$\br{N\kappa}^{-1}=1$ indicate the amount left to extrapolate. The
three curves are essentially occupying the same distance, especially
those at the smaller separations $\kappa=0.001$ and $\kappa=0.0001$,
and for easier comparison we have included a magnification of the
curves. The nearly overlapping curves indicate that the convergence
of the matrix inversion is determined mainly by the product
$N\kappa$.

Now, having confirmed the assumption, the heuristic extrapolation
method works as follows:

\begin{enumerate}

\item Order the considered $\kappa$-values as $\kappa_0 > \kappa_1
> \kappa_2 > \ldots$, with $N, f_0\brn{N}, \tilde{\mathcal{C}}$ and
$h$ indexed analogously.

\item Starting with $\kappa_0$, for which we already can assume a
fairly good convergence by using a moderate number of expansion
functions (see Figure \ref{fig:hnp}), and using the power law
(\ref{eq:NR:1}), with a truncation number $N_{0}$, we obtain the
extrapolation $\hat{\mathcal{C}}_0$. This is taken as the starting
value, $\tilde{\mathcal{C}}_0=\hat{\mathcal{C}}_0$, for our improved
algorithm, which from now proceeds repetitively:

\item Increment the index (denoted $i$). Use equation
(\ref{eq:NR:2}) to find the $h$-function from the previous step:
\begin{align}
h_{i-1}\brn{N_{i-1}\kappa_{i-1}}  = f_{0,{i-1}}\brn{N_{i-1}} -
\tilde{\mathcal{C}}_{i-1}.\label{eq:NR:2b}
\end{align}
Let $n_{i-1}$ be a truncation number fulfilling
\begin{align}
N_i \kappa_i = n_{i-1} \kappa_{i-1}. \label{eq:NR:3}
\end{align}
Assuming that $h_{i}=h_{i-1}$, it follows from (\ref{eq:NR:2}) and
(\ref{eq:NR:3}) that
\begin{align}
f_{0,i}\brn{N_i} - \tilde{\mathcal{C}}_i = f_{0,{i-1}}\brn{n_{i-1}}
- \tilde{\mathcal{C}}_{i-1} =
f_{0,{i-1}}\brn{\frac{\kappa_i}{\kappa_{i-1}}N_i} -
\tilde{\mathcal{C}}_{i-1},
\end{align}
from which the extrapolated capacitance becomes
\begin{align}
\tilde{\mathcal{C}}_i  = f_{0,i}\brn{N_i} +
\tilde{\mathcal{C}}_{i-1} -
f_{0,{i-1}}\brn{\frac{\kappa_i}{\kappa_{i-1}}N_i}.  \label{eq:NR:4}
\end{align}
Note that $n_{i-1}$, determined from (\ref{eq:NR:3}), must fulfil
$n_{i-1} \leq N_{i-1}$, and if $n_{i-1}$ does not becomes an integer
the last term in (\ref{eq:NR:4}) must be evaluated by interpolation
between the adjacent integer values.

\item Repeat step 3 until the final (smallest) $\kappa$-value has been
considered.

\end{enumerate}

\subsection{Results for the capacitance}

At small separations, the first approximation to the capacitance is
the geometric capacitance $\mathcal{C}_\text{g}=\pi /(4\kappa)$. A
much better approximation is the result by Kirchhoff
\cite{Kirchhoff}:
\begin{align}
\mathcal{C}_\text{k} \approx \frac{\pi}{4\kappa} +
\frac{1}{4}\ln\brn{\frac{1}{\kappa}} +
\frac{1}{4}\brh{\ln\brn{16\pi} - 1} + o\brn{1}.\label{eq:NR:5}
\end{align}
This formula, which has been proved rigourously by Hutson
\cite{Hutson}, becomes increasingly accurate as $\kappa$ decreases.

Our results are divided into two cases. In the first case, we used a
constant value $N\kappa = 3$, and considered separations down to
$\kappa=0.0002$; the results are given in Table \ref{tabell:Ccalc1}.
Here, all extrapolations, $\hat{\mathcal{C}}$, were obtained by
fitting the power law (\ref{eq:NR:1}) to the $f_0$-values at $N,
N/2, N/3$. We also present the relative excess
$(\tilde{\mathcal{C}}-\mathcal{C}_\text{g})/\mathcal{C}_\text{g}$
over the geometric capacitance. Our results are compared with the
Kirchhoff result (\ref{eq:NR:5}) and with the numerical results in
\cite{Wintle_Kurylowicz,Carlson_Illman}.

\begin{table}[h]
\centering
\begin{tabular}{|l||r|r|r|c||r|c|c|}
  \hline
$\quad\;\kappa$   &  $N\;\;$ & $f_0\brn{N}\quad$ &
$\hat{\mathcal{C}}\qquad$ & Exc. $\mathcal{C}_\text{g}$
  & Eq.(\ref{eq:NR:5})\hspace{3mm} & Ref.\cite{Wintle_Kurylowicz} & Ref.\cite{Carlson_Illman}\\
\hline\hline
0.01    &  300  & 80.43440 & 80.43451 & 2.41 \%     & 80.42044 & 80.4342 & 80.43 \\
0.005   &  600  & 159.14169 & 159.14179 & 1.31 \%     & 159.13354 & 159.13  & 159.1 \\
0.002   & 1500  & 394.98596 & 394.98607 & 5.82 \prom  & 394.98206 & 394.87  & 395   \\
0.001   & 3000  & 787.85661 & 787.85672 & 3.13 \prom  & 787.85443 & 787.6   & 787   \\
0.0005  & 6000  & 1573.42707 & 1573.42718 & 1.67 \prom  & 1573.42588 & 1573    & \\
0.0002  & 15000 & 3929.84994 & 3929.85005 & 7.28 \pprom & 3929.84944
& 3928.9  & \\ \hline
\end{tabular}

\caption{Numerically calculated capacitances, where $f_0\brn{N}$ is
the non-extrapolated value and $\hat{\mathcal{C}}$ is the
extrapolant. The estimated true capacitance is $C=4\ep_0 a
\hat{\mathcal{C}}$.} \label{tabell:Ccalc1}
\end{table}

In the second case, we considered even smaller separations, down to
$\kappa=0.00001$; the results are presented in Table
\ref{tabell:Ccalc2}. Since limited computer memory enforced a
maximum truncation number $N=15000$, all extrapolations were
obtained with our heuristic method. As the starting value for the
repetitive extrapolation, we took the $\hat{\mathcal{C}}$-value
obtained at the smallest separation considered in Table
\ref{tabell:Ccalc1}, i.e. $\kappa_0=0.0002, \tilde{\mathcal{C}}_0 =
3929.85005$; see Section \ref{avsnitt:numeriska:extrapol2}.

\begin{table}[h]
\centering
\begin{tabular}{|l||c|r|r|c||r|c|}
  \hline
$\quad\;\kappa$   &  $N$ & $f_0\brn{N}\quad\;$ &
$\tilde{\mathcal{C}}\qquad$ & Exc. $\mathcal{C}_\text{g}$
  & Eq.(\ref{eq:NR:5})\hspace{5mm} & Ref.\cite{Wintle_Kurylowicz} \\
\hline\hline
0.0001 & 15000 & 7857.01294 & 7857.01378 & 3.86 \pprom & 7857.01355 & 7855.9  \\
0.00005 & 15000 & 15711.16055 & 15711.16855 & 2.04 \pprom & 15711.16847 &  \\
0.00002 & 15000 & 39273.25402 & 39273.34241 & 0.87 \pprom & 39273.34244 &  \\
0.00001 & 15000 & 78543.05664 & 78543.42381 & 0.46 \pprom & 78543.42390 &  \\
\hline
\end{tabular}

\caption{Numerically calculated capacitances, where $f_0\brn{N}$ is
the non-extrapolated value and $\tilde{\mathcal{C}}$ is the
extrapolant. } \label{tabell:Ccalc2}
\end{table}

\begin{figure}[h] \centering
\psfrag{p}{$\kappa$} \psfrag{C}{$\brb{\tilde{C}-C_\text{k}}$}
\includegraphics{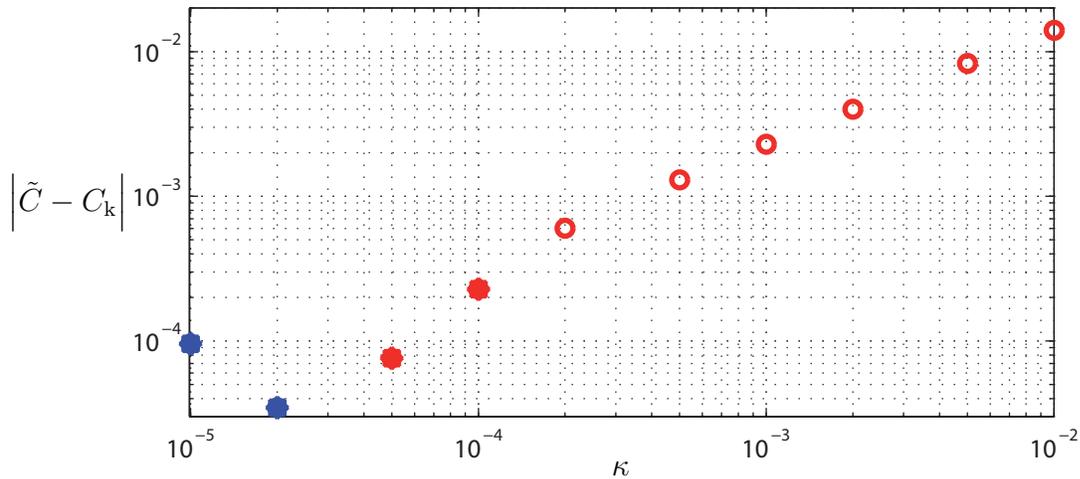}
\caption{The magnitude of the difference between our extrapolated
results and the Kirchhoff results from (\ref{eq:NR:5}), as function
of the relative separation $\kappa$. Red and blue colors denote
positive and negative differences, respectively. The circles are the
results from Table \ref{tabell:Ccalc1} and the dots are the results
from Table \ref{tabell:Ccalc2}.} \label{fig:diff_C_Ck}
\end{figure}

At the largest separation, $\kappa=0.01$, our result agrees better
with the reference numerical results than with (\ref{eq:NR:5}), but
for the smaller separations our results are more close to
(\ref{eq:NR:5}). In Figure \ref{fig:diff_C_Ck}, we have plotted the
difference between our results and the Kirchhoff result, which in
effect is our numerical approximation of the rest term in Equation
(\ref{eq:NR:5}). For $0.0002 \leq \kappa \leq 0.01$ the difference
is positive and decreases with $\kappa$ with the approximate
behavior $\propto\kappa^{0.8}$, but when reaching the smallest
$\kappa$-values the difference turns negative and increases in
magnitude. This cannot by itself be taken as an error, since
(\ref{eq:NR:5}) is not the exact solution, neither a lower bound,
but a solution that becomes increasingly accurate as $\kappa$
decreases.

We should also mention the result by Ignatowsky
\cite{Ignatowsky,Wintle_Kurylowicz}, which differs from Kirchhoff's
result in that the constant term is replaced by $\br{\ln 8 -
1/2}/4$, thereby falling below the Kirchhoff result by the amount of
approximately $0.33447$. P\'olya and Szegö \cite{Polya+Szegoe} have
shown that Ignatowsky's result is a sharp lower bound for the
capacitance. Thus, we see in Table \ref{tabell:Ccalc2} that at
$\kappa=0.00001$ our non-extrapolated result is below this lower
bound, which is an indication that convergence has not been reached
due to an insufficient number of expansion functions. However, the
extrapolated result is above the sharp lower bound.

\subsection{On the accuracy of the results}

Our numerical simulations have shown that for a fixed $\kappa$ the
unextrapolated capacitance $f_0\brn{N}$ increases with $N$, and from
Tables \ref{tabell:interpolering} and \ref{tabell:Ccalc1} it appears
that when using the power law extrapolation formula (\ref{eq:NR:1})
the extrapolated value $\hat{\mathcal{C}}$ decreases with $N$. If we
in Table \ref{tabell:interpolering} use the
$\hat{\mathcal{C}}$-values at $N\kappa=1$ as references, the
extrapolations obtained at $N\kappa=0.2$ overestimate the references
with about one third of the total amount of extrapolation.
Similarly, if we in Table \ref{tabell:Ccalc1} use the
$\hat{\mathcal{C}}$-values at $N\kappa=3$ as references, the
extrapolations obtained at $N\kappa=1$ (in Table
\ref{tabell:interpolering}) overestimate the references with about
one third of the total amount of extrapolation. Applying this rule
to the values in Table \ref{tabell:Ccalc1}, we conclude that the
true values are approximately $4\cdot 10^{-5}$ below the
extrapolated value.

Using the subsequent repeated extrapolation, the error obtained at
$\kappa=0.0002$ is propagated to the lower $\kappa$-values. If the
extrapolation scheme was ideal, i.e. if the $h$-function in
(\ref{eq:NR:2}) was the same regardless of $\kappa$, no further
contributions to the error would occur. To get a very rough estimate
of the cumulated errors, we can study the similarities between the
curves in Figure \ref{fig:hnp}. Given the excellent agreement
between the curves at the smaller $\kappa$-values, a pessimistic
estimate is that the cumulated error is within 1/10 of the amount of
extrapolation. Applying this estimate to the values given in Table
\ref{tabell:Ccalc2} we obtain the estimate of the maximum error,
denoted $\Delta \mathcal{C}$, acquired in each step of the
interpolation scheme. The results, rounded upward to one significant
digit, are presented in Table \ref{tabell:errcalc}. Since we
typically have to let $N\kappa$ decrease with $\kappa$, the most
significant contribution to the accumulated error at a certain
$\kappa$-value is acquired in the last step of the extrapolation
scheme.
\begin{table}[h]
\centering
\begin{tabular}{|c||c|c|c|c|c|}
\hline
$\kappa$  &  0.0002 & 0.0001 & 0.00005 & 0.00002 & 0.00001 \\
\hline $\Delta \mathcal{C}$ & $4\cdot 10^{-5}$ & $8\cdot 10^{-5}$ &
$8\cdot 10^{-4}$ &
$9\cdot 10^{-3}$ & $4\cdot 10^{-2}$  \\
\hline
\end{tabular}
\caption{Estimated order of the cumulated error during each step of
the heuristic extrapolation scheme. } \label{tabell:errcalc}
\end{table}

\section{Conclusions}

\label{avsnitt:diskussion}

By expanding the kernel in the Love equation into a Fourier cosine
series, with the coefficients expressed analytically in terms of
Sine and Cosine integrals, we have increased the accuracy of the
expansion, making it possible to use considerably larger truncation
numbers $N$ than in previous studies. In this way, we have improved
the numerical values for the capacitance at small plate separations
$\kappa$.

The present method enable us to consider smaller plate separation
distances than has been considered before, while maintaining a high
accuracy. Numerical tests indicated that the degree of convergence
is determined by the product $N\kappa$. Hence, for this method of
calculating the capacitance it becomes practically impossible to
accommodate for a decreasing plate separation by a corresponding
increase of the number of expansion functions; cf. Figure
\ref{fig:hnp}. To compensate for this problem, we have developed a
heuristic extrapolation scheme that uses the information obtained at
intermediate separations, where the convergence is good, to improve
the convergence at small separations.

At larger separations, $\kappa$, than those considered in our study,
convergence is obtained for smaller numbers, $N$, of expansion
functions. In such cases, one does not need to use the analytical
results in Section \ref{avsnitt:integraler} for the integrals, since
they can instead be calculated accurately by means of numerical
methods \cite{Carlson_Illman,Wintle_Kurylowicz}. In fact, for large
$\kappa$-values the explicit expressions in Section
\ref{avsnitt:integraler} are unsuitable for numerical evaluation.
This can be seen by observing that from (\ref{eq:PF:8}) and
(\ref{eq:PF:4}) it follows that for any $\kappa
> 0$ the matrix elements fulfil the relation
\begin{align}
\brb{K_{mn}} \leq 2 K_{0,0} < 2  \label{eq:discuss:1}
\end{align}
and that when we tested the algorithm for separations $1 \leq \kappa
\leq 10$ we encountered spurious large matrix elements violating the
condition (\ref{eq:discuss:1}). The reason is that when $N \kappa
\gg 1$, the individual terms in the expressions given in Section
\ref{avsnitt:integraler} exhibit an exponential growth $\propto
\e^{N\kappa}$ and the condition (\ref{eq:discuss:1}) is met by
taking the differences between such very large terms, but
numerically this kind of evaluation leads to cancelation effects.

\end{document}